**The DARPA Quantum Network**

Chip Elliott

Principal Engineer, BBN Technologies

10 Moulton Street, Cambridge, MA 02138

celliott@bbn.com

**Introduction**

It now seems likely that Quantum Key Distribution (QKD) techniques can provide practical building blocks for highly secure networks, and in fact may offer valuable cryptographic services, such as unbounded secrecy lifetimes, that can be difficult to achieve by other techniques. Unfortunately, however, QKD's impressive claims for information assurance have been to date at least partly offset by a variety of limitations. For example, traditional QKD is distance limited, can only be used across a single physical channel (e.g. freespace or telecommunications fiber, but not both in series due to frequency propagation and modulation issues), and is vulnerable to disruptions such as fiber cuts because it relies on single points of failure.

To a surprising extent, however, these limitations can be mitigated or even completely removed by building QKD *networks* instead of the traditional, stand-alone QKD links. Accordingly, a team from BBN Technologies, Boston University, and Harvard University has recently built and begun to operate the world's first Quantum Key Distribution network under DARPA sponsorship[1].

The DARPA Quantum Network became fully operational on October 23, 2003 in BBN's laboratories, and in June 2004 was fielded through dark fiber underneath the streets of Cambridge, Mass., to link our campuses with non-stop quantum cryptography, twenty-four hours per day. It is the world's first quantum cryptography network and indeed probably the first metro-area QKD deployment in continuous operation. As of December 2004, it consists of six QKD nodes. Four are used in BBN-built, interoperable weak-coherent QKD systems running at a 5 MHz pulse rate through telecommunications fiber, and inter-connected via a photonic switch. Two are NIST-built electronics for a high speed free-space QKD system. All run BBN's full suite of production-quality QKD protocols. In the near future, we plan to add four more quantum cryptographic nodes based on a variety of physical phenomena, and start testing the resulting network against sophisticated attacks.

This chapter introduces the DARPA Quantum Network as it currently exists and briefly outlines our plans for the near future. We first describe the motivation for our work and define the basic principles of a quantum cryptographic *network* (which may be composed of a number of QKD systems with relays and/or photonic switches). We then discuss the specifics of our current weak-coherent QKD network, including its QKD links, photonic switches for "untrusted" networks, and key relay protocols for "trusted" networks. We conclude with future plans and our acknowledgements.

---

[1] The opinions expressed in this article are those of the author alone, and do not necessarily reflect the views of the United States Department of Defense, DARPA, or the United States Air Force.

**Current Status of the DARPA Quantum Network**

Figure 1 displays a fiber diagram of the DARPA Quantum Network's buildout through Cambridge, Mass., as of December 2004. The network consists of two weak-coherent BB84 transmitters (Alice, Anna), two compatible receivers (Bob, Boris), and a 2x2 switch that can couple any transmitter to any receiver under program control. Alice, Bob, and the switch are in BBN's laboratory; Anna is at Harvard; and Boris is at Boston University. The fiber strands linking Alice, Bob, and the switch are several meters long. The Harvard-BBN strand is approximately 10 km. The BU-BBN strand is approximately 19 km. Thus the Harvard-BU path, through the switch at BBN, is approximately 29 km. All strands are standard SMF-28 telecommunications fiber.

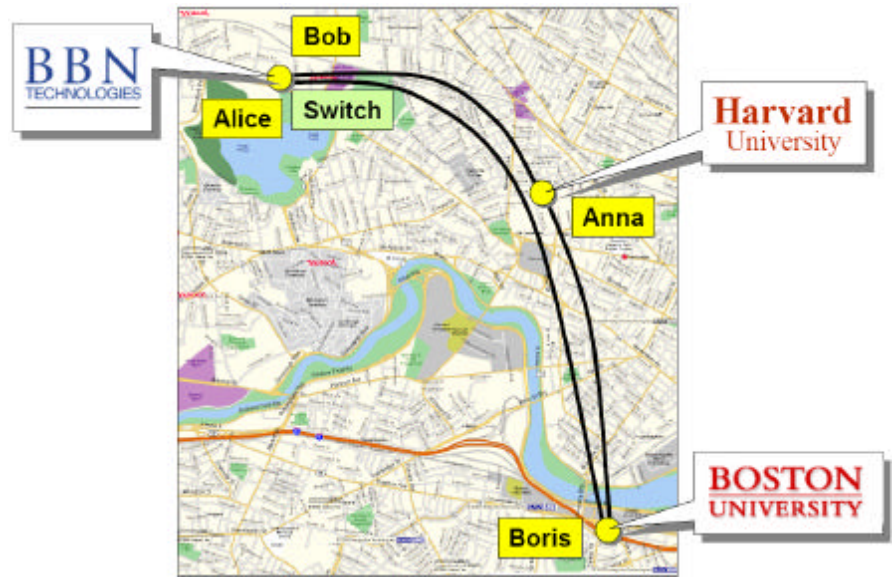

Figure 1. The Metro-Fiber Portions of the DARPA Quantum Network.

Anna's mean photon number is 0.5 at present, with the Anna-Bob link delivering about 1,000 privacy-amplified secret bits/second at an average 3% Quantum Bit Error Rate (QBER). At present the DARPA Quantum Network cannot support fractional mean photon numbers to Boris at BU, due to high attenuation in fiber segments across the Boston University campus and relatively inefficient detectors in Boris. (BBN-BU attenuation is approximately 11.5 dB). Thus the network currently operates at a mean photon number of 1.0 on the BBN-BU link, in order to continuously exercise all parts of the system, even though the resultant secret key yield is zero. In the near future, fiber splices and perhaps detector upgrades should allow operation to BU with mean photon numbers of 0.5.

The DARPA Quantum Network also contains Ali and Baba, the electronics subsystems for a high-speed freespace QKD system designed and built by the National Institute of Standards and Technology (NIST). Ali and Baba run the BBN QKD protocols, and are linked into the overall network by key relay between Ali and Alice. It further contains two new entanglement-based nodes named Alex and Bard, built jointly by BU and BBN, but these nodes are not yet fully operational.

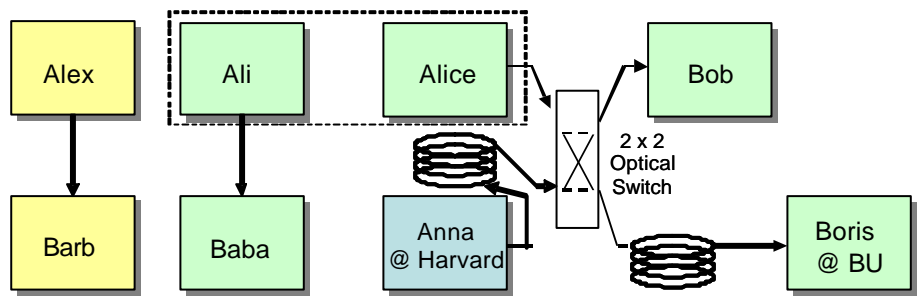

Figure 2. Connectivity Schematic of the DARPA Quantum Network.

**Motivation for the DARPA Quantum Network**

QKD provides a technique for two distinct devices to come to agreement upon a shared random sequence of classical bits, with a very low probability that other devices (eavesdroppers) will be able to make successful inferences as to those bits' values. Such sequences may then used as secret keys for encoding and decoding messages between the two devices. In short, it is a cryptographic key distribution technique[2]. Although QKD is an interesting and potentially very useful technique for key distribution, it is not the only one – human couriers and algorithmic "one-way" functions such as Diffie-Hellman come immediately to mind – and thus it is important to gauge QKD's strengths across a number of important goals for key distribution systems in general. Table 1 provides such an assessment of "classic" QKD techniques; see [1] for a more extended treatment of this subject.

Table 1. Assessment of "Classic" Quantum Cryptography.

| Important Goals for a Cryptographic Key Distribution System | QKD Strengths and Weaknesses |
| --- | --- |
| Protection of Keys | QKD offers significant advantages in this regard and indeed this is the main reason for interest in QKD. Assuming that QKD |

---

[2] Strictly speaking, it is a means for coming to agreement upon a shared key, rather than a way to distribute a key, but we follow conventional QKD terminology in this chapter.

| | |
|---|---|
| | techniques are properly embedded into an overall secure system, they can provide automatic distribution of keys that may offer security superior to that of its competitors. |
| Authentication | QKD does not in itself provide authentication. Current strategies for authentication in QKD systems include prepositioning of secret keys at the distant device, to be used in hash-based authentication schemes, or hybrid QKD-public key techniques. |
| Robustness | This critical property has not traditionally been taken into account by the QKD community. Since keying material is essential for secure communications, it is extremely important that the flow of keying material not be disrupted, whether by accident or by the deliberate acts of an adversary (i.e. by denial of service). Here QKD has provided a highly fragile service to date since QKD techniques have implicitly been employed along a single point-to-point link. |
| Distance- and Location- | This feature is notably lacking in QKD, which requires the two entities to have a direct and unencumbered path for photons |
| Independence | between them, and which can only operate for a few tens of kilometers through fiber. |
| Resistance to Traffic Analysis | QKD in general has had a rather weak approach since most setups have assumed dedicated, point-to-point QKD links between communicating entities which has thus clearly laid out a map of the underlying key distribution relationships. |

It can be seen that "classic" QKD, i.e., QKD performed by sending a single quantum entity directly from source to destination, has areas of weakness mixed with its strengths. As important guidelines of our overall research agenda, we are working to strengthen QKD's performance in these weaker areas. A surprising number of these weaknesses, as it turns out, can be removed by weaving individual QKD links into an overall QKD Network such as the DARPA Quantum Network.

**What is a QKD Network?**

Figure 3 depicts a typical, stand-alone QKD system in highly schematic form[3]. In this example, Alice contains both a photon source and a modulator; in this case, Alice

---

[3] In fact, it depicts our 'Mark 2' weak-coherent link but corresponding forms of high-level schematics

employs an attenuated laser and Mach-Zehnder interferometer. Bob contains another modulator and photon detectors, specifically a twin Mach-Zehnder interferometer and cooled InGaAs APDs. Here the channel between Alice and Bob is a standard telecommunications fiber.

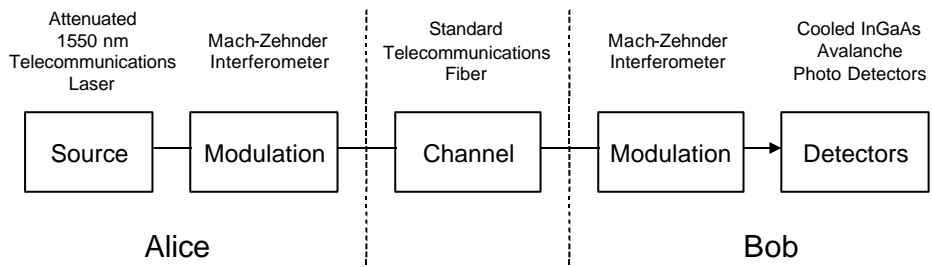

Figure 3. High-Level Schematic of an Exemplary QKD Link.

In the network context, this system can be viewed as a single, isolated QKD "link" that allows Alice and Bob to agree upon shared cryptographic key material. In our terminology, both Alice and Bob are "QKD endpoints," as are other cryptographic stations based on QKD technology, e.g., the Alice or Bob used in plug-and-play systems, entanglement-based links with Ekert protocols, and so forth.

---

can be drawn for any quantum cryptographic system.

Figure 4 shows how a number of such QKD endpoints and links may be woven together into an overall QKD *network*. Here we see that one Alice/Bob pair (A1, B1) is directly connected via a fiber strand, while another (A2, B2) is connected by a freespace channel. Other pairs are connected via fibers and photonic switches, e.g., either A1 or A3 may be connected to B3 depending on the setting of the switch between them. Multiple QKD endpoints, e.g., (B1, A2), may also be grouped into a "key relay device," whose purpose is explained below.

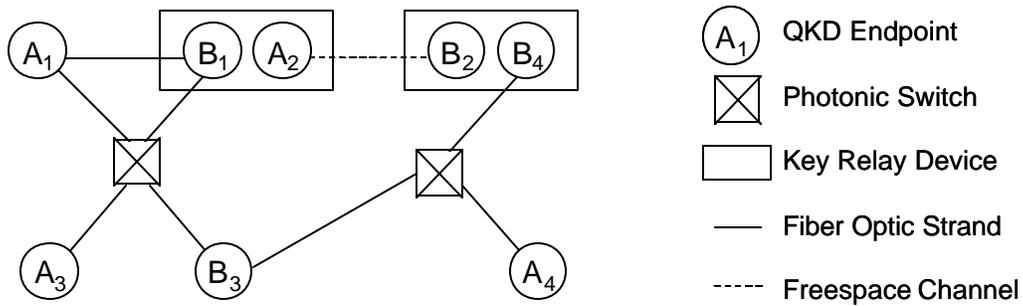

Figure 4. High-Level Schematic of an Exemplary QKD Network.

By proper use of QKD networking protocols, a node such as A1 may agree upon key material not just with its direct neighbors (B1 or B3), but indeed with nodes many hops away through the key distribution network. For example, it could agree upon shared keys with B4, through intermediaries of B1, A2, and B4. Perhaps more surprisingly, it could also agree upon shared keys with other transmitters even though neither transmitter can detect the other's photons! Thus two transmitting nodes such A1 and A3 can agree upon shared keys in a quantum cryptographic network – provided that they rely on a trusted relay, such as B1, to act as a middleman in this process.

**Photonic Switching for "Untrusted Networks."** Untrusted networks employ unamplified, all-optical paths through the network mesh of fibers, photonic switches, and endpoints. Thus a photon from its source QKD endpoint proceeds, without measurement, from switch to switch across the optical QKD network until it reaches the destination endpoint at which point it is detected. The (A1, B3) path in Figure 3 provides an example, though in general a path may transit multiple photonic switches.

Untrusted QKD networks support truly end-to-end key distribution – QKD endpoints need not share any secrets with the key distribution network or its operators. This feature could be extremely important for highly secure networks. Unfortunately, though, untrusted switches cannot extend the geographic reach of a QKD network. In fact, they may significantly reduce it since each switch adds at least a fractional dB insertion loss along the photonic path. In addition, it will also prove difficult in practice to employ a variety of transmission media within an untrusted network, since a single frequency or modulation technique may not work well along a composite path that includes both fiber and freespace links.

**Key Relay for "Trusted Networks."** After a set of QKD nodes have established pairwise agreed-to keys along an end-to-end path between two QKD endpoints – e.g., (A1, A4) in Figure 3 – they may employ these key pairs to securely relay a key "hop by hop" from one endpoint to the other, being onetime-pad encrypted and decrypted with each pairwise key as it proceeds from one relay to the next. In this approach, the end-to-end key will appear in the clear within the relays' memories proper, but will always be encrypted when passing across a link.

Key relays bring important benefits but are not a panacea. They can extend the geographic reach of a network secured by quantum cryptography, since wide-area networks can be created by a series of point-to-point links bridged by active relays. Furthermore, links can employ heterogeneous transmission media, i.e., some may be through fiber while others are freespace. Thus in theory such a network could provide fully global coverage. However, QKD key relays must be *trusted*. Since keying material and – directly or indirectly – message traffic are available in the clear in the relays' memories, these relays must not fall into an adversary's hands. They need to be in physically secured locations and perhaps guarded if the traffic is truly important. In addition, all users in the system must trust the network (and the network's operators) with all keys to their message traffic.

**The Major Benefits of QKD Networks.** Table 2 below summarizes the major benefits that QKD networks bring to traditional, stand-alone QKD links.

Table 2. Major Benefits of Quantum Cryptographic Networks.

| Benefit | Discussion |
|---|---|
| Longer Distances | QKD key relay can easily extend the geographic reach of quantum cryptography. As one example, quantum cryptography could be performed through telecommunications fiber across a distance of 500 km by interposing 4 relays between the QKD endpoints, with a span of 100 km fiber between each relay node. |
| Heterogenous Channels | QKD key relay can mediate between links based on different physical principles, e.g., between freespace and fiber links, or even between links based on entanglement and those based on weak laser pulses. This allows one to "stitch together" large networks from links that have been optimized for different criteria. |
| Greater Robustness | QKD networks lessen the chance that an adversary could disable the key distribution process, whether by active eavesdropping or simply by cutting a fiber. When a given point-to-point QKD link |

|  | within the network fails – e.g. by fiber cut or too much eavesdropping or noise – that link may be abandoned and another used instead. Thus QKD networks can be engineered to be resilient even in the face of active eavesdropping, fiber cuts, equipment failures, or other denial-of-service attacks. A QKD network can be engineered with as much redundancy as desired simply by adding more links and relays to the mesh. |
|---|---|
| Cost Savings | QKD networks can greatly reduce the cost of large-scale interconnectivity of private enclaves by reducing the required (N x N-1) / 2 point-to-point links to as few as N links in the case of a simple star topology for the key distribution network. |

**BBN's 'Mark 2' Weak Coherent Systems**

This section describes the four fiber-based QKD systems currently running in the DARPA Quantum Network. All became operational in October 2003; we call these 'Mark 2' systems because they replaced our first-generation system which started continuous operation in December 2002. These links were inspired by a pioneering Los Alamos system [2].

Each Mark 2 link employs a highly attenuated telecommunications laser (hence the term "weak coherent") at 1550.12 nm, phase modulation via unbalanced Mach-Zehnder interferometers, and cooled avalanche photo detectors (APDs). Most Mark 2 electronics are implemented by discrete components such as pulse generators, though it would not be difficult to integrate all electronics onto a small custom board. Figure 5 depicts Anna and Boris, our first rack-mounted versions of the Mark 2 hardware, before their deployment into wiring closets at Harvard and Boston University as part of the metro network.

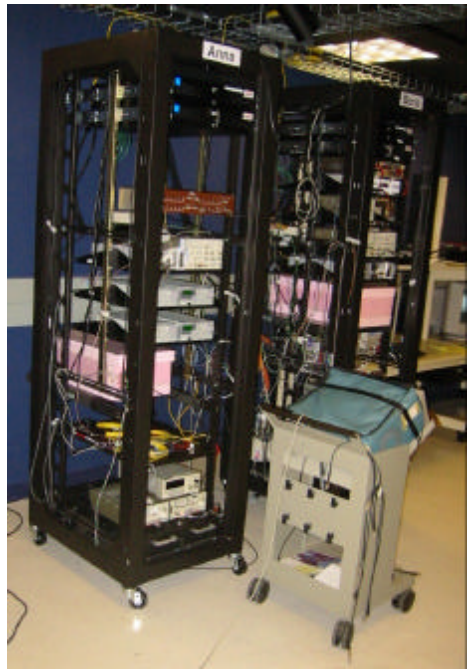

Figure 5. BBN's Mark 2 Weak-Coherent Transmitter and Receiver (Anna and Boris).

Two aspects of the Mark 2 Weak-Coherent link are fundamental to its overall operation and performance. It is important to understand these fundamental points, why they occur, and their implications on the overall system.

First, this link is designed to run through telecommunications fiber as widely deployed today. Thus we have chosen to transmit dim pulses in the 1550.12 nm window for maximal distance through this fiber. At present, these dim pulses can be best detected by certain kinds of commercial APDs cooled to approximately –40 degrees Centrigrade. These cooled detectors form one of the most important bottlenecks in the overall link performance, as they require on the order of 10 usec to recover between detection events. The overall link has been designed to run at up to 5 MHz transmit rate but with a dead-time circuit to disable the APD after a detection event in order to accommodate this recovery interval and detector after-pulsing.

Second, the Mark 2 Weak-Coherent link employs an attenuated telecommunications laser as its source of dim ("single photon") pulses. This is certainly the easiest kind of source to build. However such attenuated, weak-coherent sources have been shown to be vulnerable to at least theoretical forms of attack from Eve by Brassard et al [3]. Such attacks are generally termed "Photon Number Splitting" (PNS) attacks. For experimental purposes, we sometimes run the so-called "Geneva sifting protocol," or SARG, [4] which may provide protection against such attacks, even in systems employing attenuated laser pulses. In addition, our forthcoming entangled link will employ a completely different type of source, namely, the BU entangled source. (If possible, we may attempt to build an Eve that can actively attack the attenuated pulses used in our Mark 2 Weak-Coherent link and perform laboratory demonstrations of this heretofore theoretical form of attack.)

Figure 6 highlights the major features of our Mark 2 weak-coherent link. As shown, the transmitter at Alice sends data by means of very highly attenuated laser pulses at 1550.12 nm. Each pulse passes through a Mach-Zehnder interferometer at Alice and is randomly modulated to one of four phases, thus encoding both a basis and a value in that photon's self interference. The receiver at Bob contains another Mach-Zehnder interferometer, randomly set to one of two phases in order to select a basis for demodulation. The received single photons pass through Bob's interferometer to strike one of the two cooled detectors and hence to present a received value. Alice also transmits bright pulses at 1550.92 nm, multiplexed over the same fiber, to send timing and framing information to Bob.

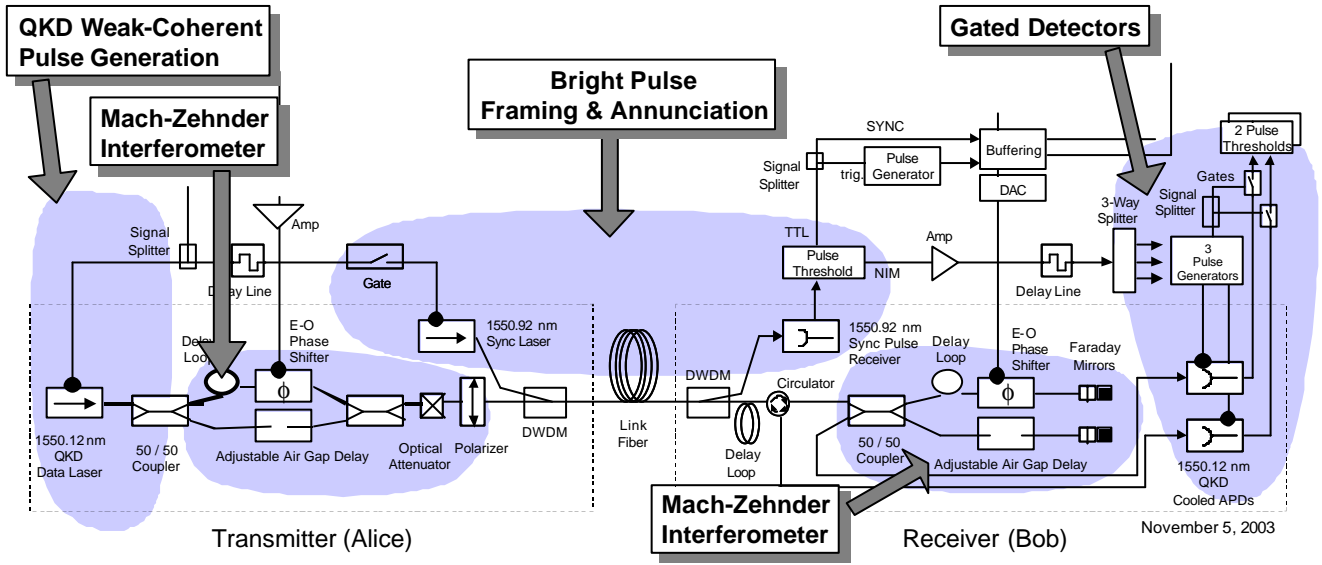

Figure 6. Functional Decomposition of the Mark 2 Weak-Coherent QKD Link.

Alice provides the clock source for both transmitter and receiver. All clocking in this system ultimately derives from a single trigger supplied from the transmitter suite. The rising edge of this signal triggers a pulse generator whose output is split into two pulses: one drives the 1550.12 nm QKD data laser to create data pulses and the other the drives the 1550.92 nm sync laser through a gate and delay line. The delay line provides a stable time relationship between the data and synch pulses, and is chosen so that the sync pulse is transmitted about 20 ns after its associated data pulse.

The data pulse passes through the unbalanced Mach-Zehnder interferometer where one arm applies phase shift modulation to the pulse. A D-to-A converter drives an electro-optic modulator with an analog voltage that produces the Basis and Value phase shifts clocked from the transmitter electronics.  In the other arm an adjustable air gap delay line allows fine-tuning of the interferometer differential delay.  After exiting the interferometer the data pulse is attenuated to achieve the requisite mean photon number.  A polarizer then removes mistimed replicas of the data pulse that may have been generated by misaligned polarization-maintaining components in the interferometer.  At the transmitter output the data pulse is combined with the sync pulse in a DWDM optical multiplexing filter.

At the receiver the sync and data pulses are separated with a DWDM filter and the sync pulse is detected with a PIN-FET receiver.  This signal is shaped in a pulse thresholding circuit that produces two outputs: a 100 ns TTL-level clock signal sent to the receiver electronics and a 4 ns NIM-level APD gate-timing pulse that triggers the APD gate-pulse generators and the pulse generator driving the APD output line gates.  The output line gates are timed to pass only the demodulated data signal from the APDs and block noise due to spurious pulse reflections.  An adjustable delay line in the NIM pulse interconnection allows fine-tuning of APD gate-pulse timing.

The data pulse passes through a fiber delay loop to adjust its timing with respect to the sync pulse and then through a circulator that is the input to the interferometer demodulation circuit.  This interferometer is a folded version of conventional Mach-Zehnder design and is independent of the input polarization to accommodate the uncontrolled incident polarization at the receiver. Faraday mirrors at the ends of the unequal-length arms reflect light in such a way that the polarization of the light returning to the beam splitter is the same for each arm, producing interference with high visibility [5]. The Basis is clocked out of the receiver electronics and applied to the electro-optic modulator through a D-to-A converter to produce a phase shift of either 0 or $\pi/2$. A pair of cooled APDs, biased above avalanche breakdown only during the time a data photon is expected to arrive, detect the interferometer outputs, one from the beam splitter and the other from the circulator. After gating to select only the data pulse, the APD signals are shaped by threshold detectors and passed as "0" or "1" to the receiver electronics.

A phase-correcting feedback signal, derived by the receiver from training frames sent by the transmitter, is used to maintain phase stability between the transmitter and receiver interferometers as path lengths change with temperature and stress. This phase-correcting signal is applied to the receiver interferometer electro-optic modulator through the transmitter electronics. Phase correction is also necessary when a transmitter and receiver first connect during a start-up or switching operation to obtain the phase-matched condition needed for low quantum bit error rate. See [6] for a discussion of BBN's algorithms for automatic path-length control.

It is by now well-known that certain conventional InGaAs APDs can be operated in the single-photon regime if properly cooled and gated. Like many other research teams, we have selected Epitaxx EPM 239 AA APDs for our detectors. Even with this special treatment, they suffer considerably from low Quantum Efficiency (QE), relatively high dark noise, and serious after-pulsing problems. Even so, they provide adequate performance for a 5-MHz pulse rate quantum cryptography system. Since custom cooling and electronics are required, we designed and built our own cooler package to maintain the InGaAs APDs at the requisite operating temperatures. We had two key goals in mind for the cooler package: it should be able to operate reliably and repeatably over a wide range of temperatures down to –80 C, to enable exploration of detector behavior of over a range of operating conditions; and it should be suitable for prototype deployment and thus should not require human intervention on a regular basis, e.g., to refill liquid nitrogen reservoirs.

Figure 7 shows a schematic of the core housing, which is a vacuum-pumped chamber containing a pair of InGaAs detectors brought to operating temperature by two Peltier thermo-electric coolers (TECs).

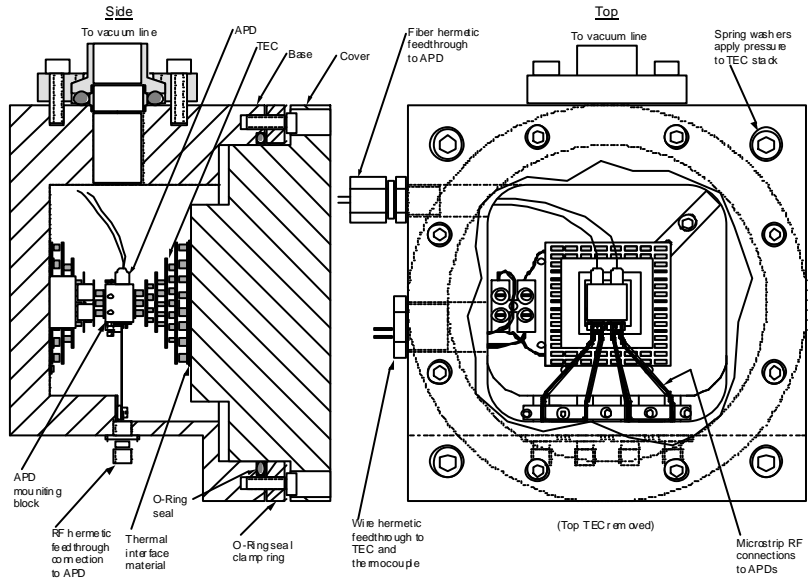

Figure 7. Side and Top Cutaway Views of DARPA Thermo-Electric Cooler Package.

The housing itself is machined aluminum, consisting of a base container plus a removable lid. The lid fits snugly with an O-ring so that a vacuum can be maintained in the inner chamber. A number of holes pass through the base container to allow hermetic feedthrough of fibers and electrical connections. A large connection leads to the vacuum line. At the center of the chamber rests a cooled block of copper with holes drilled out for two detectors. The detectors are inserted into this block, with fibers fed to the outside through a hermetic seal. Microstrip RF connections lead from the side of the container to the detectors. These two detectors are the two 1550.12 nm QKD Cooled APDs, i.e., one is D0 and the other is D1. Each detector has its own fiber lead through which come the dim pulses for that detector. Each detector also has a set of microstrip RF connectors by which the bias voltage may be applied and the detector output led to electronics outside the chamber.

The block of copper rests between two 5-stage Peltier coolers, also known as Thermo-Electric Coolers (TECs). We choose two coolers of this size to ensure that we had adequate margin for whatever range of temperatures we wanted to explore; two or three stages would suffice for routine operation. The chamber also contains a thermocouple to measure its current temperature. Electrical leads for the coolers and thermocouple pass through hermetic feedthroughs to the outside equipment.

**BBN QKD Protocols**

Although a detailed discussion of BBN's QKD protocols is well beyond the scope of this chapter, quantum cryptographic systems contain a surprising amount of sophisticated software. It has been our observation that the optics in quantum cryptography is perhaps the easiest part; the electronics are more difficult than the optics; and for a real, functional system, the software is harder than the electronics.

Figure 8 illustrates our software architecture in a high-level form. Here we see that the QKD protocols have been integrated into a Unix operating system and provide key material to its indigenous Internet Key Exchange (IKE) daemon for use in cryptographically protecting Internet traffic via standard IPsec protocols and algorithms. See [7] for a more detailed discussion of this implementation and how QKD interacts with IKE and IPsec.

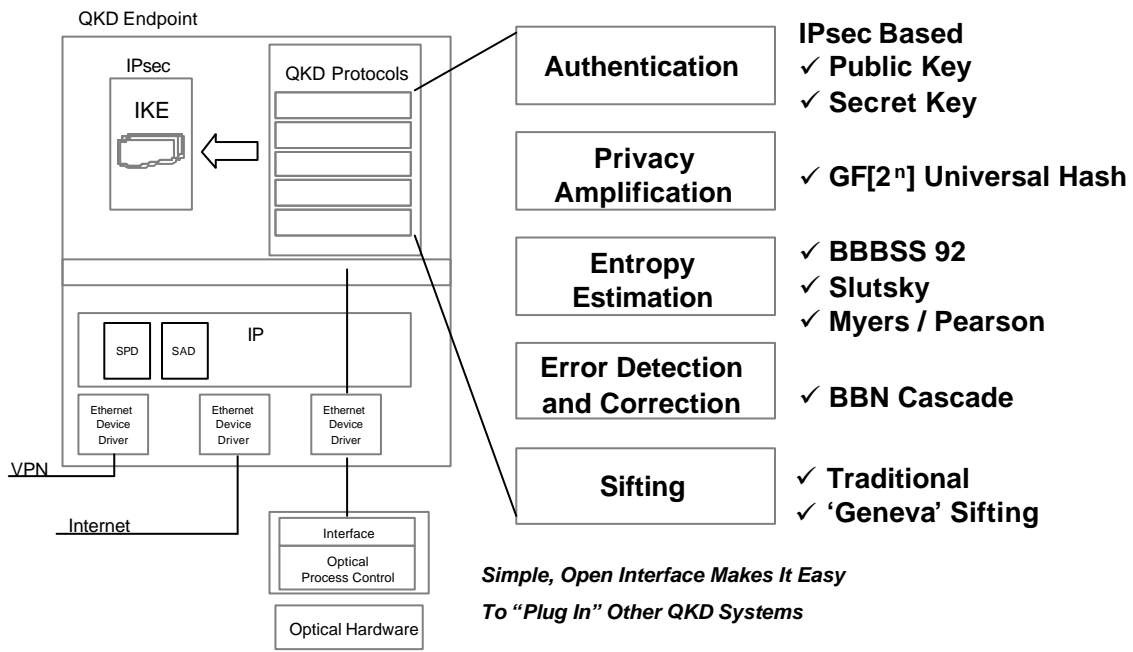

Figure 8. The BBN QKD Protocol Suite in Context.

BBN's QKD protocol stack is an industrial-strength implementation written in the C programming language for ready portability to embedded real-time systems. At present all protocol control messages are conveyed in IP datagrams so that control traffic can be conveyed via an internet. However, the control messages could be ported to use other forms of communications quite easily, e.g., ATM networks or dedicated channels.

Two aspects of BBN's QKD protocol stack deserve special mention. First, it implements a complete suite of QKD protocols. In fact, it implements multiple "plug compatible" versions of some functions as shown in Figure 7; for instance, it provides both the traditional sifting protocol and the newer "Geneva" style sifting [4][4]. It also provides a choice of entropy estimation functions. We expect to add additional options and variants as they are developed. Second, BBN's QKD protocols have been carefully designed to make it as easy as possible to plug in other QKD systems, i.e., to facilitate the introduction of QKD links from other research teams into the overall DARPA Quantum Network.

**Photonic Switching for Untrusted Networks**

The DARPA Quantum Network currently consists of two transmitters, Alice and Anna, and two compatible receivers, Bob and Boris, interconnected through their key transmission link by a 2 x 2 optical switch. In this configuration, either transmitter can directly negotiate a mutual key with either receiver. The switch must be optically passive so that the quantum state of the photons that encode key bits is not disturbed.

---

[4] For some time, we ran traditional sifting during week days and Geneva sifting over the weekends to gain realistic experience with both.

Figure 9 depicts the fiber-based portion of the current network diagram. Here all four QKD endpoints are connected through a conventional 2x2 optical switch. At one switch position, Alice is connected to Bob, and Anna to Boris. At the other, Alice is connected to Boris, and Anna to Bob. At present the switch controller changes this connectivity on a periodic basis, e.g., every 15 minutes. Immediately after the switch setting is changed, the receivers autonomously discover that they are receiving photons from a new transmitter, and realign their Mach-Zehnder interferometers to match the transmitter's interferomenter. Then they begin to develop new key material by performing the BBN QKD protocols with this new transmitter.

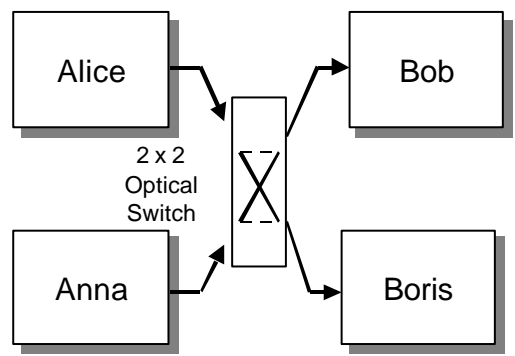

Figure 9. Current Topology of the DARPA Quantum Network.

The switch chosen for this network is a standard telecommunications facilities switch that operates by moving reflective elements that change the internal light path to produce either a BAR or CROSS connection. It is operated by applying a TTL-level pulse to either the BAR or CROSS pin for 20 ms and latches in the activated position. Switching time is 8 ms and optical loss is <1 dB. Figure 10 shows a photograph of the switch mounted on a PC board with the electrical interconnects on the left and optical interconnects on the right.

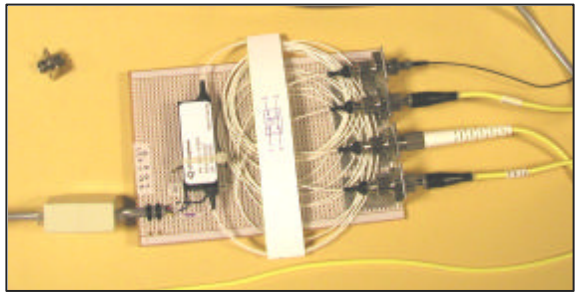

Figure 10. The 2x2 optical switch mounted on a PC board.

**BBN Key Relay Protocols for Trusted Networks**

When two QKD endpoints do not have a direct or photonically-switched QKD link between them, but there is a path between them over QKD links through trusted relays, novel BBN-designed networking protocols allow them to agree upon on shared QKD bits. They do so by choosing a path through the network, creating a new random number R, and essentially sending R one-time-pad encrypted across each link. We call this process "key relay," and the resultant network a "trusted network" since the chief characteristic of this scheme is that the secrecy of the key depends not just on the endpoints being trustworthy; the intermediate nodes must also be trusted.

The BBN key relay protocols have been continuously operational in the DARPA Quantum Network since October 2003. In fact, they run continuously in our network through Cambridge, and allow Alice to build up a reservoir of shared key material with Anna, even though both entities are transmitters, via a trusted relay at Bob or Boris. Similarly Bob and Boris continuously build up shared key material via trusted relays at Alice or Anna.

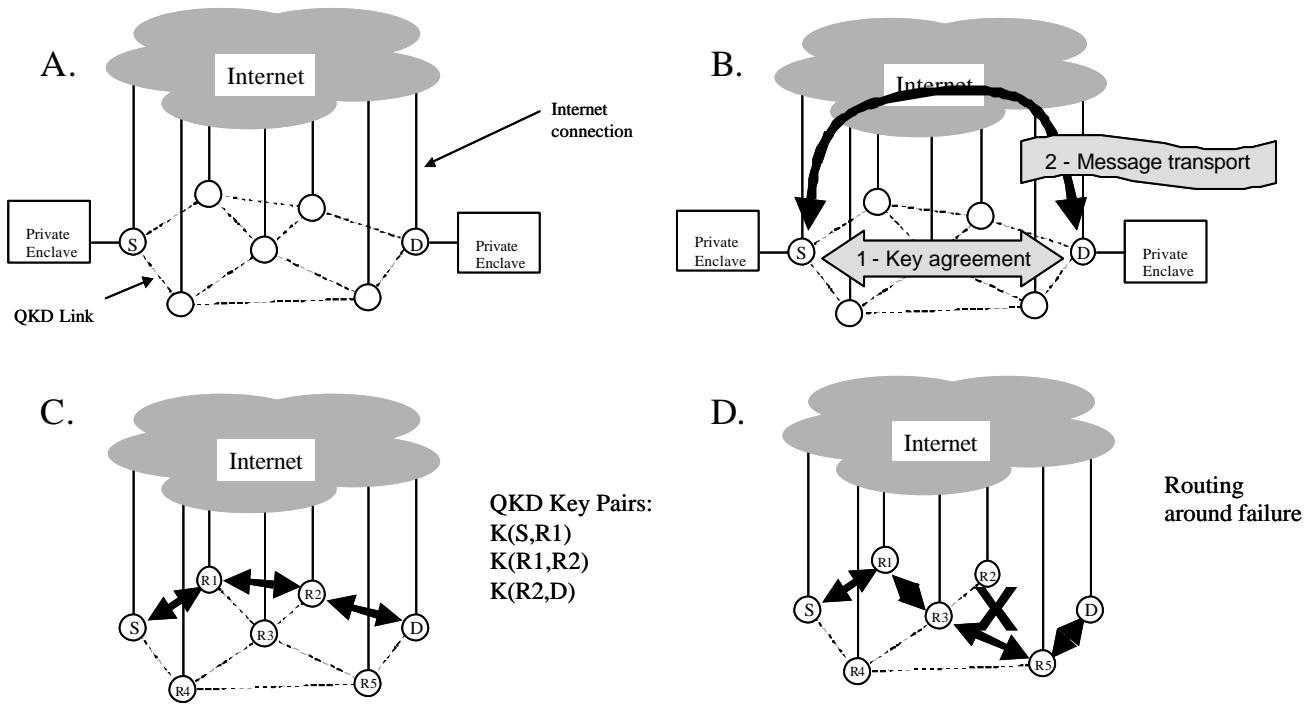

Figure 11. Major Aspects of the Key Relay Protocols.

The four main aspects of key relay are illustrated in Figure 11. Figure 11-A makes it clear that a key relay network is parallel to an overarching network conveying communication messages and control traffic such as the QKD protocols. Here the Internet is the communications network, each link underneath it is a separate QKD link, and circular nodes are key relay stations. In Figure 11-B, one particular source QKD endpoint (S) wishes to agree upon key material with a far-away destination QKD endpoint (D). Since both endpoints, S and D, are connected to a ubiquitous communications network, they can perform QKD protocols in order to derive key material, and once they have agreed upon these keys, they can then use the Internet to communicate between themselves securely.

Figure 11-C shows a path used for key relay from S to D, as darkened lines across the key relay network, and the resultant pairs of QKD key material at the right. One QKD-derived key is shared between S and R1; this key is denoted K(S,R1). Likewise K(R1,R2) denotes the key pairwise shared between relay nodes R1 and R2, and so forth. Once all these pairwise keys are in place, S and D can easily derive their own end-to-end shared secret key by key relay. One obvious means is for node S to create a new random number R, protected this number R by K(S,R1), and transmit the result to R1. Node R1 can then decrypt this message to obtain R itself, and re-encrypt it by K(R1, R2) to send it onwards to R2, who can in turn repeat the process, and so forth, until it has been relayed all the way to D. At this point, both S and D know the same secret random sequence, R, and can use this shared value as key material.

Finally, Figure 11-D shows the BBN key relay protocols can automatically discover failures along the key relay path – whether due to cut fiber or eavesdropping – and route the key material around these failures.

**Future Plans**

Our near-term plans call for augmenting the DARPA Quantum Network with four new QKD nodes: one pair based on entangled photons in fiber, and the other on polarized photons in a freespace channel.

- The entangled link's optical subsystem has been designed and shaken down by Boston University, and is now resident at BBN's laboratory. All electronics and software have been built. Once the entangled system is fully operational, we will tie it via key relay into the overall DARPA Quantum Network.
- The freespace link will be based on polarization modulation of faint laser pulses at visible wavelengths. The transmitter will contain 4 lasers, one for each polarization basis and value, which pulse according to externally-supplied random signals; the receiver will perform passive random splitting via a 50/50 coupler. This link will also be woven into the DARPA Quantum Network when operational.

**Summary**

The DARPA Quantum Network has married a variety of QKD techniques to well-established Internet technology in order to build a secure key distribution system that can be employed in conjunction with the public Internet or, more likely, with private networks based on the Internet Protocol suite. Such private networks are currently in widespread use around the world with customers who desire secure and private communications, e.g., financial institutions, governmental organizations, militaries, and so forth.

The DARPA Quantum Network has been in continuous operation in BBN's laboratory since October 2003, and since June 2004 through dark fiber linking the Harvard, Boston University, and BBN campuses. It currently consists of four interoperable QKD nodes designed for use through telecommunications fiber, and a passive photonic switch that interconnects them; a high-speed free-space system designed and built by NIST; and a full suite of production-quality QKD protocols running on all nodes. Key material derived from these systems is integrated into the Internet security protocols (IPsec) to protect user traffic.

**Acknowledgements**

We are deeply indebted to Dr. Mike Foster (DARPA IPTO) and Dr. Don Nicholson (Air Force Research Laboratory) who are the sponsor and agent, respectively, for this research project. We also thank Dr. Carl Williams and his team at NIST for their generous, long-term loan of their QKD system. This paper reflects highly collaborative work between the project members. Of these, particular credit is due to Profs. Alexander